\documentclass{elsart}

\usepackage{graphicx}

\usepackage{amssymb}
\usepackage{amsmath}

\begin{document}

\begin{frontmatter}

\title{Charge Fractionalization in nonchiral Luttinger systems}

\author[Yale]{Karyn Le Hur}, 
\author[Harvard]{Bertrand I. Halperin}, 
\author[Harvard]{Amir Yacoby}

\address[Yale]{Department of Physics, Yale University, New Haven, CT 06520, USA}
\address[Harvard]{Department of Physics, Harvard University, Cambridge MA, 02138, USA}

\begin{abstract}
One-dimensional metals, such as quantum wires or carbon nanotubes, can 
carry charge in arbitrary units, smaller or larger than a single electron charge. However, according to
Luttinger theory, which describes the low-energy excitations of such systems, when a single electron is injected by tunneling into the middle of such a wire, it will tend to break up into separate charge pulses, moving in opposite directions, which carry  definite fractions $f$ and $(1-f)$ of the electron charge, determined by a parameter $g$ that measures the strength of charge interactions in the wire.
(The injected electron will also produce a spin excitation, which will travel at a different velocity than the charge excitations.) Observing charge fractionalization
physics in an experiment is a challenge in those (nonchiral) low-dimensional
systems which are adiabatically coupled to Fermi liquid leads. We theoretically discuss a first
important step towards the observation of charge fractionalization in quantum wires  
based on momentum-resolved tunneling and multi-terminal geometries, and explain the recent
experimental results of H. Steinberg {\it et al.}, Nature Physics {\bf 4}, 116 (2008).
\end{abstract}

\end{frontmatter}


\section{Introduction}

Systems of low dimensions have provided special opportunities, challenges, and fascination for condensed matter physicists \cite{Giamarchi}. Issues of long-range order, dimensional crossover, and instabilities are all significant in such systems. Many aspects of the physics have been studied, and many are now understood. The electronic properties of low-dimensional systems such as quantum wires or carbon nanotubes are in many cases well described by the Tomonaga-Luttinger theory \cite{Luttinger,Matthis,Haldane} which predicts spectacular effects of electron-electron interactions such as {\em electron fractionalization}: an injected electron will necessarily break up into separate charge and spin excitations, which will travel at different velocities.  The evidence of spin-charge separation in Luttinger systems has been obtained via the non-linear tunneling conductance between parallel wires in a transverse magnetic field \cite{Auslaender}. This provides a direct spectroscopic probe of spin-charge separation which gives a similar information to an ideal photoemission experiment. Luttinger liquid behavior has also been observed through energy dependent local tunneling \cite{Bockrath} and power-law tunneling lineshapes \cite{Auslaender1}.

Another remarkable effect predicted by Luttinger liquid theory is {\em charge fractionalization}: the extra charge  produced by an electron tunneling into the middle of a uniform Luttinger liquid will  break  up into pieces, moving in opposite directions, which will carry definite fractions  $f$ and $(1-f)$ of the electron charge, determined by a parameter $g$ (the Luttinger parameter) that measures the strength of charge interactions in the wire. This  fractional charge is not just a statistical average, but is predicted to be a definite property of the quantum mechanical state,  subject in principle to verification by repeated measurements after a single tunneling event \cite{Pham,Karyn,Ines,Imura,TMartin}. 

It should be emphasized that charge fractionalization in a (gapless) Luttinger liquid is very different from the charge fractionalization in  a two-dimensional fractional quantized Hall system.  For example, if the quantized Hall system is in a Laughlin state \cite{Laughlin}  at Landau-level filling fraction  $\nu=1/3$,
a  quasiparticle  with  minimal fractional charge $e/3$ is an absolutely stable excitation. If it is far from a boundary and far from other excitations, the charge cannot further subdivide,  and it will be separated by an energy gap from other states with multiple quasiparticles and quasiholes, which  could carry the same total charge. In a nonchiral Luttinger liquid, it is possible to create left and right moving charge excitations with arbitrary charge, at arbitrarily small energy\footnote{See the discussion in Appendix B.4, below.}. The definite charges $fe$ and $(1-f)e$ are most properly understood as a property of the electron injection process, rather than as properties of elementary excitations in the liquid itself. 

The edge of a quantized Hall system is a peculiar, one-dimensional metal, with  gapless charged excitations, which has many properties in common with the ordinary nonchiral Luttinger liquid \cite{Wen}.  The edges of a quantized Hall state are chiral, however, so that, in most cases, charge can flow in only one direction a long any given edge. By charge conservation, an electron injected from the outside into one edge of a macroscopic quantized Hall system must propagate along that edge, with charge $e$.   However, if  there is a narrow constriction in a strip containing electrons in a fractional quantized Hall state, there can be tunneling of charge from one edge to the other, and this charge need not be quantized in units of $e$.  If the quantized Hall system is in a Laughlin state at  $\nu=1/3$, and if the matrix elements for  tunneling across the constriction are sufficiently weak, then charge will typically be transferred from one edge to the other in unit of $e/3$ \cite{Fisher}. However, in the more general case, where tunneling between the edges is not weak, it is not clear  that charge is necessarily transferred in multiples of $e/3$.  Thus for the chiral Luttinger liquid, just as for the ordinary Luttinger liquid, it may be most precise to regard charge fractionalization in a chiral Luttinger liquid as a consequence of the charge injection process, rather than as a fundamental property of the (chiral) Luttinger liquid itself. However, this issue requires further investigation.

Observing charge fractionalization physics in an experiment has generally been a considerable challenge, even for fractional quantized Hall systems. However, the existence of isolated charge of magnitude $e/3$ in the bulk of a fractional quantized Hall state at $\nu=1/3$ has recently been confirmed by  direct charge sensing with a single electron transistor \cite{Martin}.  The predicted fractional charge for tunneling across a constriction in the weak backscattering regime has been  confirmed by low-frequency shot-noise measurements \cite{shot-noise} in selected quantized Hall systems. A number of other experiments have also given evidence for fractional charge \cite{Su}. In particular, an $e/3$-quasiparticle interferometer has been envisioned theoretically \cite{Chamon}  and realized experimentally recently \cite{Goldman}.

While for fractional quantum Hall edge states, the counterpropagating modes are spatially separated, in quantum wires or carbon nanotubes, the nonchiral modes are confined to the same spatial channel, and cannot be contacted individually. Their chemical potentials renormalize in a non-trivial manner when adiabatically coupled to metallic leads, making charge fractionalization phenomenoa difficult to observe. The dc two-terminal conductance with ideal contacts is universal and independent of interactions \cite{Ines,Maslov,Ponomarenko}. Furthermore, low-frequency shot-noise measurements in an ideal wire would only reveal the physics of the Fermi-liquid contacts \cite{Nagaosa,Grabert,TMartin1}. A straightforward transposition of the results obtained for the chiral edge system therefore proves difficult. 

Very recently, we have envisioned a three-terminal geometry where unidirectional electrons are injected from the bulk of a wire and the resulting current at drains located on both sides is measured, to detect charge fractionalization \cite{Amir}. More precisely, using momentum conservation in the tunneling process between two wires one injects unidirectional electrons to the bulk of a wire, with charge fractionalization resulting in currents detected on both sides of the injection region. We have suggested that the ratio of these currents together with a two-terminal reference measurement then allows to extract the extent of charge fractionalization. In this paper, we flesh out the theoretical concepts and demonstrate that such a three-terminal geometry reveals charge fractionalization in
accordance with the recent experimental results \cite{Amir}. In particular, we introduce a novel universal ratio which allows to reveal the charge fractionalization mechanism in nonchiral Luttinger liquids.

\section{Charge fractionalization from unidirectional injection}

\subsection{Chiral basis}

Firstly, we introduce the chiral fields of nonchiral Luttinger liquids \cite{Giamarchi,Pham,Karyn,Imura,Fisher,Ines2}. It is indeed convenient to distinguish the charge excitations propagating to the left $(-)$ from the charge excitations propagating to the right $(+)$. 

The charge sector of a (single-mode) quantum wire is described by the Luttinger Hamiltonian (the electron spectrum is assumed to be perfectly linear),
\begin{equation}
H = \frac{u \hbar}{2}\int_0^L dx \left[\frac{1}{g}\left(\frac{\partial \phi}{\partial x}\right)^2 + g  \left(\frac{\partial \theta}{\partial x}\right)^2\right],
\end{equation}
where $\phi$ is the charge mode and $\theta$ the superfluid phase satisfying:
\begin{equation}
[\theta(x),\partial_y \phi(y)] = i\delta(x-y).
\end{equation}
Moreover, we identify $\partial_t^2\phi = u^2 \partial_x^2 \phi$, and similarly for $\theta$; this represents  (plasmon) waves moving at the velocity $u$. It should be noted that the chiral densities must satisfy 
$\rho_{\pm}=F(x\mp ut)$, where $F$ is arbitrary, since the charge sector of the Luttinger liquid is (only) controlled by the plasmon velocity $u$. In fact, one can find $\rho_{\pm}$ by writing the ``conservation'' laws $\delta \rho = (\rho_+ + \rho_-) =
-\sqrt{\frac{2}{\pi}}\partial_x \phi$ and $j = e u(\rho_+ - \rho_-) = (ug) e \sqrt{\frac{2}{\pi}}\partial_x \theta$.
We have used the definitions of the fluctuations in the charge (electron) density $\delta\rho = \rho-\rho_0$, where $\rho_0$ is the mean electron density, and of the current density in a Luttinger liquid for a spinful Luttinger liquid. The chiral densities are thus defined as \cite{Pham}:
\begin{equation}
\rho_{\pm} = \frac{1}{2} \sqrt{\frac{2}{\pi}}\left[-\partial_x \phi \pm g\partial_x \theta\right].
\end{equation}
The Luttinger Hamiltonian thus can be rewritten as:
\begin{equation}
H =  \int_0^L dx \left(\frac{u \hbar \pi}{2 g}\left[\rho_+^2 +\rho_-^2\right]\right).
\end{equation}
One can check that $[\rho_+(x),\rho_-(y)]=0$, which ensures that the charge Hamiltonian is now properly decomposed into a left-moving and a right-moving part. From the
chiral Hamiltonian and the commutation relations,
\begin{equation}
\left[\rho_{\pm}(x),\rho_{\pm}(y)\right] = \mp \frac{ig}{\pi}\partial_x\delta(x-y),
\end{equation}
we verify that $(u\partial_x\pm \partial_t)\rho_{\pm} =0$,
which ensures that $\rho_{\pm}=F(x\mp ut)$.

It is also convenient to introduce the chiral chemical potentials associated to the chiral modes, $\frac{\partial H}{\partial \rho_{\pm}} = e(Y,W)$. We deduce,
\begin{equation}
e(Y,W) = \frac{u\pi\hbar}{g}\rho_{\pm} = \frac{u h}{2g}\rho_{\pm}.
\end{equation}
Hence, the chiral currents in a quantum wire are defined as,
\begin{equation}
I_{\pm} = e\rho_{\pm} u = \frac{2e^2 g}{h}(Y,W).
\end{equation}

\subsection{Counterpropagating currents}

Now, to elucidate the concept of charge fractionalization in a quantum wire embodied by a Luttinger theory, let us inject an extra electron with a well-defined momentum at one Fermi point. 

In a geometry comprising parallel wires, this can be performed through the application of a transverse magnetic field $B$ producing a large momentum boost\footnote{The momentum boost is 
gauge-invariant. Using the Landau gauge, this momentum boost directly stems from the unidirectional vector potential which is perpendicular to the wires and to the applied magnetic field. Using the symmetric gauge and making the appropriate gauge transformation such that the kinetic term of electrons (along the wires) is gauge independent  essentially produces the same result.} $q_B = 2\pi Bd/\phi_0$ --- $d$ denotes the distance between the two wires and $\phi_0$ is the flux quantum --- parallel to the wire when an electron tunnels from the upper probe (wire) to the quantum wire. The momentum boost is independent of energy. An ideal situation to probe charged excitations in a Luttinger liquid is when the magnetic field obeys $k_{F2} + k_F = q_B = 2\pi Bd/\phi_0>0$ \cite{Amir}, where $k_{F2}$ and $k_F$ are the Fermi wavectors of the upper one-dimensional probe and of the wire, respectively; see Fig. 1. This is necessary to ensure that electrons are injected in the lower wire at one Fermi point only, say $-k_F$. 

The first manner to admit charge fractionalization is to write down a tunneling Hamiltonian between the quantum wire and a one-dimensional (1D) upper probe  which is long (compared to $1/k_F$) to conserve momentum during the tunneling process. In the case of Fig. 1, the tunneling Hamiltonian reads\footnote{We decompose the electron operator as $\Psi_{2\alpha}(x) = e^{i k_F x} \Psi_{+2\alpha}(x)+ e^{-i k_F x} \Psi_{-2\alpha}(x)$.},
\begin{eqnarray}
H_{t_2} &=& \int dx\  \sum_{\alpha=\uparrow,\downarrow} \left(t_2 \Psi^{\dagger}_{-\alpha}(x)\Psi_{+ 2\alpha}(x) +h.c.\right) \\ \nonumber
&=& \lim_{q\rightarrow 0} \int dx \sum_{\alpha=\uparrow,\downarrow} \left(t_2 e^{-iqx} 
\Psi^{\dagger}_{-\alpha}(x)\Psi_{+ 2\alpha}(x) +h.c.\right),
\end{eqnarray}
\begin{figure}
\begin{center}
\includegraphics[width=5.5cm,height=3.6cm]{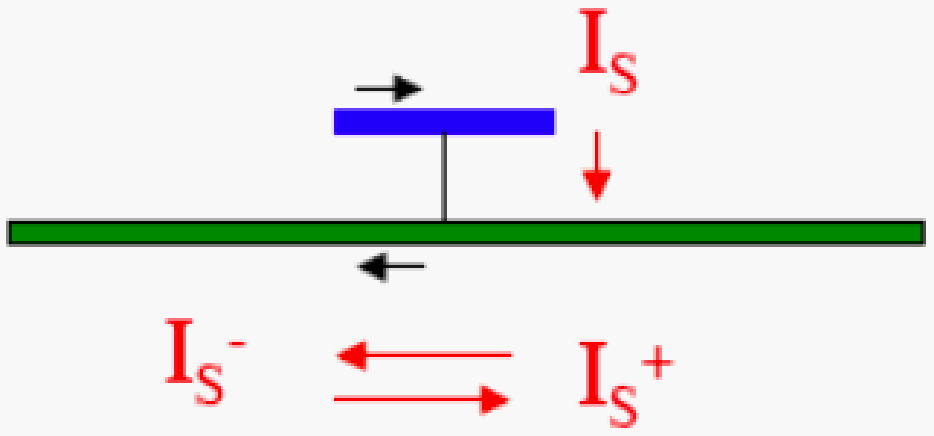}
\hskip 1cm
\includegraphics[width=4.8cm,height=3.6cm]{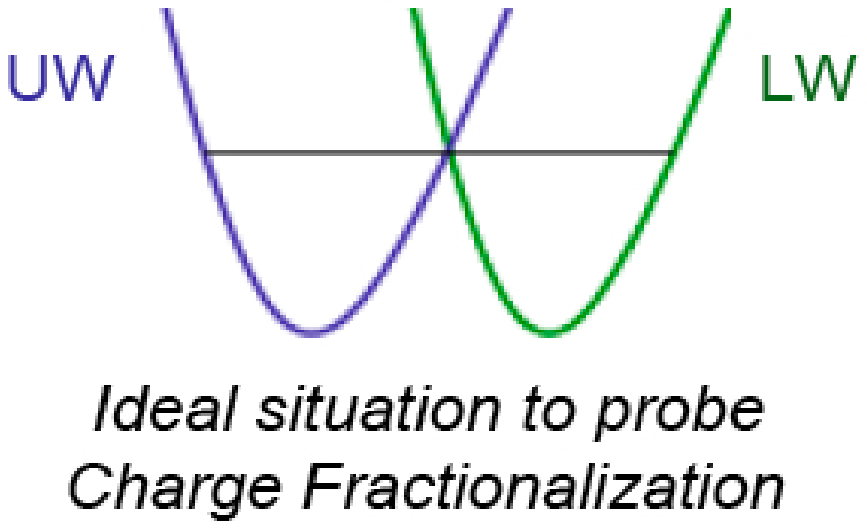}
\end{center}
\caption{Wire tunnel-coupled to an extended 1D probe (upper wire) allowing momentum-resolved tunneling; because of the uniformity of the barrier, momentum
along the wire is conserved during tunneling. A transverse magnetic field is applied so that the band structures of the two wires obey $k_{F2} + k_F = q_B = 2\pi Bd/\phi_0\gg 0$; $k_{F2}$ and $k_F$ are the Fermi wavectors of the probe and of the wire, $B$ embodies the magnetic field applied perpendicular to the plane ($q_B\approx 2k_F$ is the resulting momentum boost), $d$ is the distance between the two wires, and $\phi_0$ is the flux quantum.}
\end{figure}
where $\Psi_{+2\alpha}$ refers to a right-moving (left-moving) electron in the upper wire, $\Psi_{-\alpha}$ to a left-moving electron in the lower wire, and $t_2$ corresponds to the (electron) tunneling amplitude; a large momentum boost $q_B$ between the two wires allows to inject unidirectional electrons in the lower wire which will be essential to probe the charge fractionalization; $q=q_B-(k_F +k_{F2})\rightarrow 0$ \cite{Amir}. 

Now, let us compute the resulting current density operator in the lower wire at a point $x$ 
and time $t$. Using the Heisenberg relations, this takes the form,
\begin{eqnarray}
{j}_S(x,t) &=& \frac{ie}{\hbar} \left[ H_{t_2}, \sum_{\alpha=\uparrow,\downarrow} \Psi^{\dagger}_{-\alpha}(x,t) \Psi_{-\alpha}(x,t)\right] \\ \nonumber
&=&  -\frac{ie}{\hbar} t_2 \sum_{\alpha=\uparrow,\downarrow} \left(\Psi^{\dagger}_{-\alpha}(x,t)\Psi_{+2\alpha}(x,t) -h.c.\right).
\end{eqnarray}
The tunneling current operator reads:
\begin{equation}
I_S(t) = -\frac{ie}{\hbar} t_2 \sum_{\alpha=\uparrow,\downarrow} 
\lim_{q\rightarrow 0} \int_0^{L_F} dx\left(e^{-iqx}\Psi^{\dagger}_{-\alpha}(x,t)\Psi_{+2\alpha}(x,t) -h.c.\right).
\end{equation}
Hereafter, we will use the length $L_F$ to characterize the broadening of the spectral features in the tunneling conductance at low temperatures; one may expect that ${L}_F$ coincides with the length of the upper (shorter) wire. On the other hand, several effects not included in the Luttinger model can also broaden the electron spectral function \cite{Balents}. For example, in Ref. \cite{Amir}, electrons in the wires also interact with those in the bulk two-dimensional electron gas (electrical contact is made to the upper wires via a two-dimensional (2D) electron gas) which results in a finite 1D-2D scattering length $l_{1D-2D}\approx 6\mu m$ such that $L_F\approx l_{1D-2D}$ at low temperatures \cite{Stormer}. 

Now, we estimate the left-going current density operator in the lower wire at the same $x$ and $t$. Using the definitions of the chiral densities, we get:
\begin{equation}
{j}_S^-(x,t) = \frac{ie}{\hbar}\left[H_{t_2}, \rho_-(x,t)\right].
\end{equation}
It is now convenient to use the definition of a left-moving electron \cite{Karyn3},
\begin{equation}
\Psi^{\dagger}_{-\alpha}(x',t) = \frac{1}{\sqrt{2\pi a}}\exp 
\left[-i\sqrt{\frac{\pi}{2}}(\theta(x',t) +\phi(x',t))\right] {S}_{-\alpha}(x',t),
\end{equation}
where $a$ is the usual short-distance cutoff (mean-distance between electrons) of the Luttinger theory and ${S}_{-\alpha}$ produces a spin excitation of spin $S^z=\alpha/2=\pm 1/2$ (spinon). This
representation, which illustrates the spin-charge separation in the clean limit, allows us to show that the spin sector will not influence the charge fractionalization mechanism. Thus,
exploiting,
\begin{eqnarray}
\left[e^{-i\sqrt{\frac{\pi}{2}}\theta(x')},\partial_{x} \phi(x)\right] &=& \sqrt{\frac{\pi}{2}}\exp \left(-i\sqrt{\frac{\pi}{2}}\theta(x')\right) \delta(x-x')\\ \nonumber
\left[e^{-i\sqrt{\frac{\pi}{2}}\phi(x')},\partial_{x} \theta(x)\right] &=& \sqrt{\frac{\pi}{2}}\exp \left(-i\sqrt{\frac{\pi}{2}}\phi(x')\right) \delta(x-x'),
\end{eqnarray}
we find:
\begin{equation}
\left[ \Psi^{\dagger}_{-\alpha}(x'), \rho_-(x)\right] = -\frac{1+g}{2} \Psi^{\dagger}_{-\alpha}(x')\delta(x-x').
\end{equation}
This results in:
\begin{equation}
\label{IS-}
j_S^-(x,t) = \frac{1+g}{2} j_S = f j_S(x,t).
\end{equation}
In a similar way, we can introduce the chiral right-going current density,
\begin{equation}
\label{IS+}
j_S^+(x,t) = \frac{ie}{\hbar}\left[H_{t_2}, \rho_+(x,t)\right]= \frac{1-g}{2} j_S(x,t)=(1-f)j_S(x,t).
\end{equation}
Hence, if one injects a unidirectional electron in a Luttinger liquid, the resulting current (operator) is not unidirectional; in contrast, one can notice that the induced charged excitations propagate in both directions. This proves that the extra charge produced by an electron tunneling into the middle of a uniform wire (described by a Luttinger liquid) will break up into two pieces.

In addition, one can compute the averaged current at a point $x$ and time $t$ in the lower wire for a finite bias voltage $V_{SD}$ applied between the upper and lower wires. For the sake of clarity, calculations (based on the Keldysh approach and on linear response in $|t_2|$) are shown in Appendix A. At the left extremity of the lower wire $(x\rightarrow 0)$, we check that the current obeys,
\begin{equation}
|\langle I(x\rightarrow 0)\rangle| =  \left(\frac{1+g}{2}\right) I_S = I_S^-.
\end{equation}
Here, $I_S$ refers to the averaged injected current $\langle I_S(t)\rangle$. In the almost equilibrium limit $V_{SD}\rightarrow 0$ and at a finite temperature $T$, to lowest order in $t_2$, we find that the injected current $I_S(q\ll 1/L_F)$ obeys the precise relation:
\begin{equation}
\frac{d I_S}{d V_{SD}}(q) =  |t_2|^2 
\frac{2e^2}{h}\frac{1}{(\hbar v_F)^2}\frac{(T/\Lambda)^{\nu}}{\Gamma(\nu+1)} 
\frac{{L}_F^2}{1+q^2{L}_F^2}.
\end{equation}
The expression of the tunneling current is in agreement with Ref. \cite{Balents}; this reflects the spectral functions of the double-wire system in the regime $V_{SD}\rightarrow 0$ \cite{Karyn2,Fiete} (in the present setup, the broadening at low temperatures essentially stems from mechanisms beyond the Luttinger theory; see Sec. 2.2). Here, $\nu\geq 0$ is the tunneling exponent which obeys  $\nu=0$ for non-interacting electrons and which can be thoroughly evaluated by taking into account the Coulomb interaction between the wires; the precise expression for $\nu$ as a function of the Luttinger exponents in the lower and in the upper wires is given in Ref. \cite{Balents}. In the case of screened (Hubbard) interactions and perfectly symmetric wires, neglecting the Zeeman shift, one finds 
$\nu = -1+(g+g^{-1})/{2}$. We have also introduced the temperature $T$ assuming that $k_B T\gg eV_{SD}$ and $\Lambda$ is an ultraviolet temperature cutoff. For simplicity, we assume that the upper and lower wires have the same Fermi velocities $v_F$ which obeys $ug=v_F$. 

At the right extremity of the lower wire, we get:
\begin{equation}
\langle I(x\rightarrow L)\rangle = \left(\frac{1-g}{2}\right) I_S = I_S^+.
\end{equation}

\subsection{Charge Fractionalization: Beyond the quantum average}

In fact, the theory predicts that nonchiral Luttinger liquids allow fractional states with irrational charges which correspond to exact eigenstates of the Luttinger model \cite{Pham}; consult Appendix B.2. In addition, it should be emphasized that those anomalous charges must satisfy simple conservation laws. 

More precisely, let us inject $N$ electrons in the lower wire such that $J$ describes the
difference between the number of right-moving electrons and the number of left-moving electrons.
In general, this will produce two counterpropagating states with charge $N_+$ and $N_-$.
Since the charge states (densities) propagate at the plasmon velocity we must satisfy ${N}_+ +
{N}_- = N$ and $u(N_+ - N_-) = ug J$. In particular, the last equality stems from the precise definition of the current in a nonchiral Luttinger liquid: $\int_0^L dx j(x) = ugeJ$.
We deduce that the charges carried by the chiral excitations must take on the values \cite{Pham}:
\begin{equation}
N_{\pm} = \frac{N\pm g J}{2},
\end{equation}
where both $N$ and $J$ are integers. 

The analysis of Sec. 2.2 confirms that if one injects an electron in the lower wire, say, at $-k_F$, this will give rise to two counterpropagating pieces of charge $fe$ and $(1-f)e$, where $f=(1+g)/2$. The charge fractionalization essentially stems from the renormalization of the charge velocity in the Luttinger theory. Moreover, Eqs. (15) and (16) emphasize that those charges are not the result of a quantum average. To check this important point, one can compute the autocorrelation noise at zero frequency $S(x,x,\omega=0)$ (see Appendix A):
\begin{equation}
S(0,0,\omega=0) = fe |\langle I(x=0)\rangle|,
\end{equation}
and,
\begin{equation}
S(L,L,\omega=0) = (1-f)e \langle I(x=L)\rangle.
\end{equation}
{\it The noise diagnosis shows that  the charges found are not a quantum average}. 

\subsection{Main goal of this paper}

$\bullet$ Herein, we focus on the case where one injects an extra electron in the lower wire, say, at the left Fermi point $(N=-J=1)$. The theory predicts that this produces two counterpropagating eigenstates (chiral modes) with irrational charges $N_-=f=(1+g)/2$ and $N_+ = (1-f) = (1-g)/2$ \cite{Pham,Karyn,Karyn2}. 

As a first step, charge fractionalization can be confirmed by measuring:
\begin{equation}
\frac{I_S^ - - I_S^+}{I_S} = (2f -1) = g. 
\end{equation}
Since, one must satisfy that $I_S^+ + I_S^- = I_S$ this demonstrates Eqs. (17) and (19).
This ratio examplifies that a bare electron will give rise to two counterpropagating waves with anomalous charges $fe$ and $e(1-f)$. As a first important step towards the experimental proof of the existence of charge fractionalization in nonchiral Luttinger liquids, we show how to measure this ratio in a three-terminal geometry. The theory presented below is in accordance with the recent experimental results found in Ref. \cite{Amir}. Point-like tunneling would not be judicious to probe charge fractionalization since it results in $I_S^-=I_S^+=I_S/2$; see Appendix B.1. It should also be mentioned that the effective charges in the shot-noise formula (21) and (22) are extremally sensitive to the couplings with measuring leads \cite{Nagaosa,Grabert,TMartin1}, {\it i.e.}, the anomalous charges are difficult to observe through shot-noise measurements at low frequency. 

$\bullet$ On the other hand, for a pure current excitation, which is produced for example from weak backscattering processes mediated by an impurity ($N=0$ and $J = \pm 2$), one rather generates Laughlin ``quasiparticle-quasihole'' pairs of charge $\pm ge$ \cite{Pham,Fisher}. Those anomalous charges have eluded detection so far.

\section{Three-terminal geometry close to equilibrium}

Even though the ratio $(I_S^-  - I_S^+)/I_S$ appears as an important quantity since it evidences the charge fractionalization in non-chiral Luttinger liquids, the currents which are measured experimentally at the left (O$_1$) and right (O$_3$) contacts are not $I_S^-$ and $I_S^+$, but rather $I_L$ and $I_R$; see Fig. 2.  Within our conventions, $I_L$ and $I_R$ refer to a {\it left-going} and {\it right-going} current, respectively. On the other hand, we will show below that for realistic experimental conditions, one can extract a (first) proof of charge fractionalization through the ratio $(I_S^-  - I_S^+)/I_S$.  We focus on momentum-resolved transport in the almost equilibrium condition $V_{SD}\rightarrow 0$ (and finite temperature) where the electron spectral function shows a Lorentzian peak at $q=0$ \cite{Karyn2,Fiete}.

\subsection{Free electron analysis}

First off, some intuition can be gained by investigating the free electron problem in such a 
three-terminal geometry.  Here, all the wires are described by non-interacting electrons
and the central probe satisfies $I_S^- = I_S$ and $I_S^+=0$.

For free electrons, one may apply the Landauer-B\"uttiker approach and show that with three probes (O$_1$, O$_2$, O$_3$) one may introduce a novel universal ratio as a result of the unidirectional
injection. Here, $T_1$ and $T_3$ correspond to the (transmission) probabilities that an electron in the lower wire escapes into the upper contacts O$_1$ and O$_3$. We treat the electrons as ``classical'' particles and not worry about the phase relationships among the different paths
assuming that the length $L$ of the lower wire is larger than the phase-relaxation
length.  

Now, let us inject an electron at $-k_F$ in the lower wire from the central probe such that $I_S^-=I_S$ and $I_S^+=0$. We introduce the asymmetry parameter $A_S=(I_L-I_R)/I_S$. For free electrons, this corresponds to the probability $T^{eff}_1$ that this (left-moving) electron escapes into the left contact O$_1$ minus the probability $T^{eff}_3$ that this electron escapes into the right contact O$_3$; more precisely, $I_L=I_ST^{eff}_1$ and $I_R=I_S T^{eff}_3$. Each probability is obtained by summing over all possible paths in the lower wire. A simple calculation gives,
\begin{eqnarray}
I_L &=& I_S \left(T_1 + (1-T_1)(1-T_3)T_1 + ... + (1-T_1)^n(1-T_3)^n T_1+...\right) \\ \nonumber
&=& I_S \frac{T_1}{1-(1-T_1)(1-T_3)},
\end{eqnarray}

\begin{figure}[ht]
\begin{center}
\includegraphics[width=9.5cm,height=4.5cm]{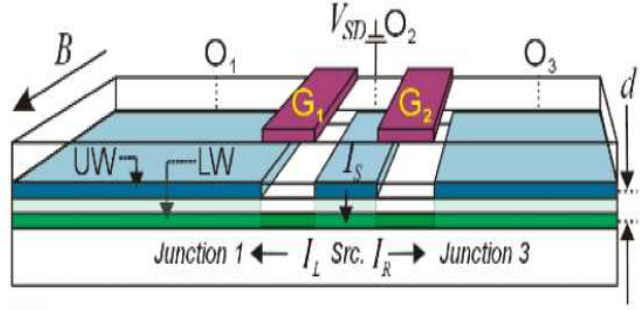}
\end{center}
\vskip -0.4cm
\caption{Three-lead geometry, where the upper wires inject unidirectional electrons, which allows to study the asymmetry parameter $A_S=(I_L-I_R)/I_S$; within our conventions, $I_L$ and $I_R$ represent the left and right currents in the system.}
\end{figure}
and to,
\begin{eqnarray}
I_R &=& I_S \left((1-T_1)T_3 + ... + (1-T_1)^{n+1}(1-T_3)^n T_3+...\right) \\ \nonumber
&=& I_S \frac{(1-T_1)T_3}{1-(1-T_1)(1-T_3)}.
\end{eqnarray}
Therefore, we deduce:
\begin{equation}
A_S = \frac{\left(T_1 - (1-T_1)T_3\right)}{1-(1-T_1)(1-T_3)}.
\end{equation}

Now, we set the (electric) potential $V_{SD}=0$ at the central junction such that $I_S=0$ and we focus on the two-terminal conductance,
\begin{equation}
G_2= \frac{I_R}{V_1-V_3}; 
\end{equation}
here, $V_1$ and $V_3$ represent the voltages at the two contacts O$_1$ and O$_3$ and the system satisfies $I_L=-I_R$ when $V_{SD}=I_S=0$. For free electrons, the two-terminal conductance can be obtained by extracting the effective transmission probability $T_{eff}$ through this ``double-barrier'' system, such that $G_2=2e^2 T_{eff}/h$. Another simple calculation gives the celebrated result:
\begin{equation}
T_{eff} = T_1 T_3\sum_{n=0}^{\infty} (1-T_1)^n (1-T_3)^n = \frac{T_1 T_3}{1-(1-T_1)(1-T_3)}.
\end{equation}
This is the expected formula in the case of two barriers in cascade and incoherent transport; in particular, one can define the ``semi-classical'' resistance ${R}_{eff}^c=h(1-T_{eff})/(2e^2 T_{eff})$ of the two barriers such that, ${R}^c_{eff} = {R}_1^c + {R}_2^c$ where ${R}_i^c = h(1-T_i)/(2e^2 T_i)$. It is appropriate to visualize $G_2^{-1}$ as,
\begin{equation}
\frac{h}{2e^2} \frac{1}{T_{eff}} = \frac{h}{2e^2} + \frac{h}{2e^2}\left(\frac{1-T_{eff}}{T_{eff}}\right),
\end{equation}
where the first term embodies the ``contact'' resistance with the leads.

From this analysis, we conclude that for symmetric couplings with the left and right upper wires, such that $T_1=T_3$, one can define a universal ratio:
\begin{equation}
\frac{A_S(2e^2/h)}{G_2} = 1.
\end{equation}
The present geometry composed of parallel wires, in principle, fullfils the necessary prerequisite $T_1=T_3$. For a very asymmetric situation where, say, $T_3\ll T_1$, in contrast, one would obtain $A_S\rightarrow 1$ whereas $G_2 h/(2e^2)\propto T_3\ll 1$.

\subsection{Charge Fractionalization and Novel Universal Ratio}

Now, we consider the case where the lower wire is realistically described by a Luttinger theory and we aim to compute the ratio $A_S(2e^2)/(h G_2)$. 

To compute this ratio, we exploit the conservation of current at the (left and right) junctions 1 and 3 of Fig. 2 and define the dimensionless parameter $\beta_1=\beta_1(q\rightarrow 0)$ ($\beta_3=\beta_3(q\rightarrow 0)$) as the transmission coefficient for a left-going (right-going) current  into the upper left (right) wire $1$ $(3)$.  In Secs.  3.3 and 3.4, we will provide a microscopic justification for the parameters $\beta_1$ and $\beta_3$. Close to perfect transmission between the upper left (right) wire and the lower wire, $\beta_i$ can be interpreted in terms of a transmission coefficient for the chiral ``quasiparticles'', whereas in the weak-tunneling limit $\beta_i$ can be precisely related to the electron tunneling amplitude $t_i$. Close to perfect transmission, we will show that $\beta_i$ formally exceeds unity;  this peculiar scenario generally arises in the context of a quantum wire perfectly coupled to Fermi liquid leads producing Andreev type reflections at the ``extremities'' of the wire \cite{Ines,Ines2}. Interactions in the upper wires will enter in the expressions of $\beta_1$ and $\beta_3$.

Since $\beta_1$ and $\beta_3$ correspond to the transmission coefficients for a left-going (right-going)
current at the junction $1$ $(3)$, when injecting a unidirectional electron in the lower wire from the central wire such that $I_S^-=f I_S = I_S(1+g)/2$ and $I_S^+=(1-f)I_S$, the currents $I_L$ and $I_R$ obey the precise forms:
\begin{eqnarray}
I_L &=& \left(I_S^- \beta_1 + I_S^+(1-\beta_3)\beta_1\right)\sum_{n=0}^{+\infty} (1-\beta_1)^n (1-\beta_3)^n \\ \nonumber
&=& \frac{\left(I_S^- \beta_1 + I_S^+(1-\beta_3)\beta_1\right)}{1-(1-\beta_1)(1-\beta_3)} \\ \nonumber
I_R &=& \left(I_S^+ \beta_3 + I_S^-(1-\beta_1)\beta_3\right)\sum_{n=0}^{+\infty} (1-\beta_1)^n (1-\beta_3)^n \\ \nonumber
&=& \frac{\left(I_S^+ \beta_3 + I_S^-(1-\beta_1)\beta_3\right)}{1-(1-\beta_1)(1-\beta_3)}.
\end{eqnarray}
Since by definition,
\begin{equation}
I_S = I_S^+ + I_S^- = I_L + I_R,
\end{equation}
then we deduce the following form of the asymmetry:
\begin{equation}
\label{As}
A_S = \left(\frac{I_S^- - I_S^+}{I_S}\right)\frac{\beta_1\beta_3}{1-(1-\beta_1)(1-\beta_3)} + \frac{\beta_1-\beta_3}{1-(1-\beta_1)(1-\beta_3)}.
\end{equation}
For free electrons, we recover Eq. (26) if we identify $\beta_i=T_i$. Assuming symmetric couplings with the left and right leads (wires), {\it i.e.}, for $\beta_1=\beta_3=\beta$, the asymmetry parameter is proportional to the fractionalization ratio:
\begin{equation}
A_S = \left(\frac{I_S^- - I_S^+}{I_S}\right)\frac{\beta}{2-\beta}.
\end{equation}
This formula can also be recovered as follows. Let us set $V_1=V_3=0$. Following Sec. 2, we introduce the chemical potentials $(W_1,Y_1)$ and $(W_3,Y_3)$ associated with the chiral densities on each side of the central junction; see Fig. 3. 

From current conservation, we get
\begin{equation}
\frac{2ge^2}{h} Y_1 = \frac{2ge^2}{h} W_1 - I_L \hskip 0.5cm  \hbox{and}\hskip 0.5cm 
\frac{2ge^2}{h} W_3 = \frac{2ge^2}{h} Y_3 - I_R,
\end{equation}
where from the definitions of the parameters $\beta_i$, one identifies:
\begin{equation}
I_L = \frac{2ge^2}{h}\beta_1 W_1\hskip 0.5cm  \hbox{and}\hskip 0.5cm 
I_R = \frac{2ge^2}{h}\beta_3 Y_3.
\end{equation}
This leads to the important equations:
\begin{equation}
Y_1 = W_1 (1-\beta_1) \hskip 0.5cm  \hbox{and}\hskip 0.5cm W_3=Y_3 (1-\beta_3).
\end{equation}
Now, we can exploit that,
\begin{equation}
I_S^- = \frac{2ge^2}{h}(W_1-W_3)  \hskip 0.5cm  \hbox{and}\hskip 0.5cm
I_S^+ = \frac{2ge^2}{h}(Y_3-Y_1),
\end{equation}
to obtain:
\begin{eqnarray}
\frac{2ge^2}{h} W_1 &=& \frac{1}{\beta_1+\beta_3-\beta_1\beta_3} \left((1-\beta_3)I_S^+ +I_S^-\right), \\ \nonumber
\frac{2ge^2}{h} Y_3 &=& \frac{1}{\beta_1 +\beta_3 -\beta_1\beta_3} \left((1-\beta_1)I_S^- +I_S^+\right).
\end{eqnarray}
\begin{figure}[ht]
\begin{center}
\includegraphics[width=12.3cm,height=2.8cm]{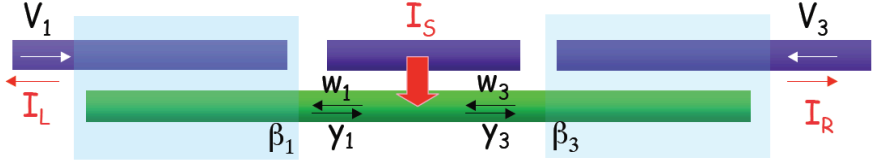}
\end{center}
\vskip -0.3cm
\caption{General Scattering Matrix formulation for the interacting case.}
\vskip 0.3cm
\end{figure}
Combining Eqs. (36) and (39) allows us to recover Eq. (\ref{As}).

Now, we shall compute the two-terminal conductance which is defined as $G_2=I_R/(V_1-V_3)$. We need to express $V_1$ and $V_3$ in terms of the relevant chemical potentials associated with the chiral densities (eigenstates). For $V_1\neq 0$ and $V_3\neq 0$, current conservation now implies the 
general equations \cite{Chklov}:
\begin{eqnarray}
\label{pot}
Y_1 &=& \gamma_1 V_1 +(1-\beta_1)W_1 \\ \nonumber
 W_3 &=& \gamma_3 V_3 +(1-\beta_3)Y_3.
\end{eqnarray}
On the other hand, $(W_1+Y_1)$ ($W_3+Y_3$) is determined by the charge density on each part of the central junction, such that in global thermal equilibrium, where there is no current (and all the bias voltages are almost equal),  $Y_1$ $(Y_3)$ and $W_1$ $(W_3)$ are equal to the electric potential $V_1$ $(V_3)$. This results in $\gamma_1=\beta_1$ and $\gamma_3=\beta_3$ \cite{Chklov}. Below, we assume the almost equilibrium limit where currents are small enough such that those equalities are always applicable. 

To compute the two-terminal conductance $G_2$ we must set $I_S=0=I_S^+=I_S^-$ such that $I_L=-I_R$ and:
\begin{equation}
W_1 = W_3 \hskip 0.5cm  \hbox{and}\hskip 0.5cm 
Y_1 = Y_3.
\end{equation}
Using Eq. (\ref{pot}), this results in:
\begin{equation}
\label{G2}
(V_1-V_3) = \left(\frac{\beta_3+\beta_1-\beta_1 \beta_3}{\beta_1 \beta_3}\right) (Y_1-W_1).
\end{equation}
Since by definition $I_R=(2e^2g/h)(Y_1-W_1)$, we find:
\begin{equation}
G_2 = \frac{2ge^2}{h} \left(\frac{\beta_1 \beta_3}{1-(1-\beta_1)(1-\beta_3)}\right).
\end{equation}
For free electrons, we check that this formula is consistent with the conductance through two scatterers in cascade; consult Eq. (28).

For $\beta_1=\beta_3=\beta$, finally we identify:
\begin{equation}
\frac{A_S(2 ge^2/h)}{G_2} =  \left(\frac{I_S^- - I_S^+}{I_S}\right) = (2f-1).
\end{equation}
\begin{figure}[ht]
\begin{center}
\includegraphics[width=8cm,height=6cm]{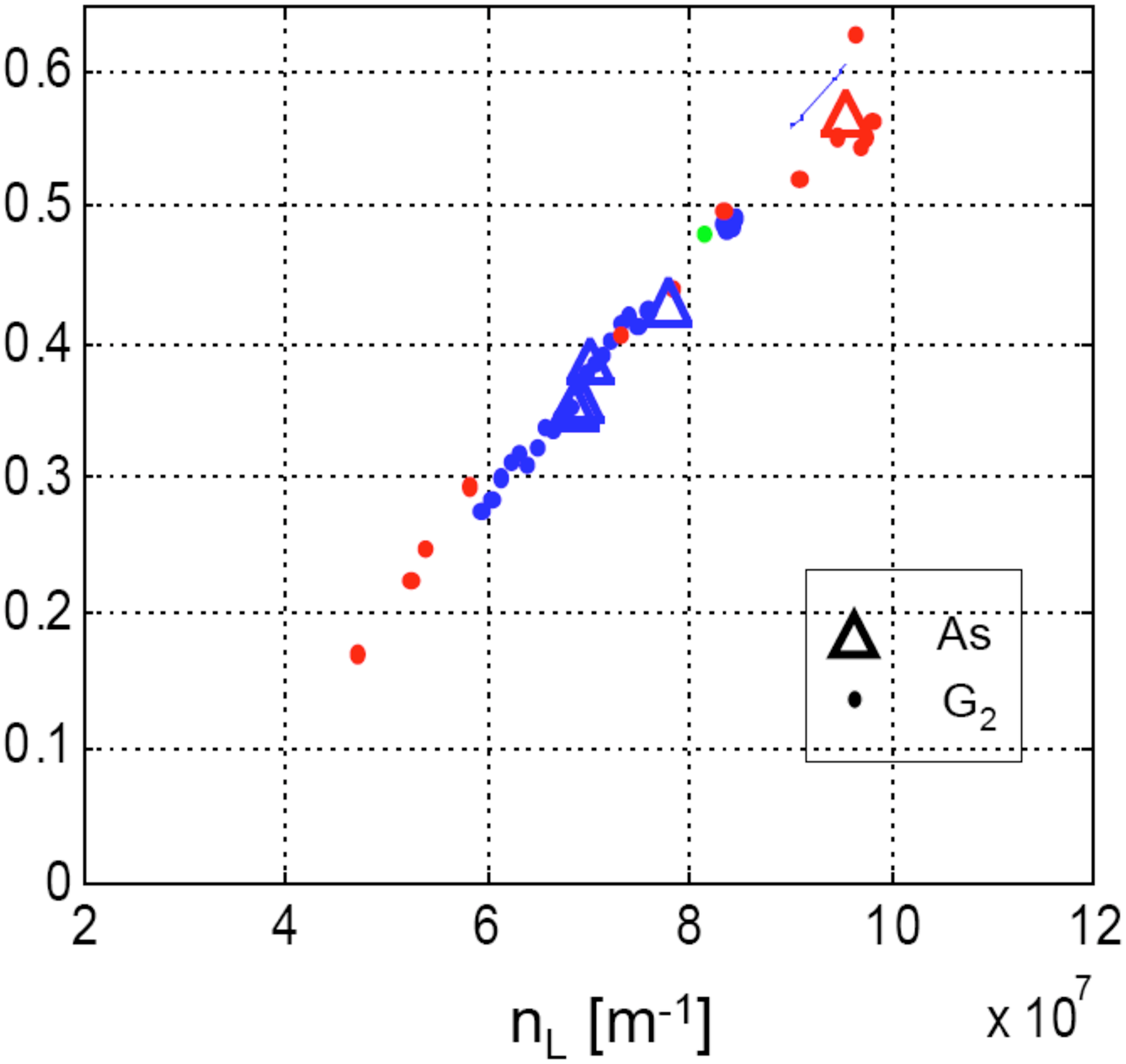}
\end{center}
\caption{Experimental results of Ref. \cite{Amir} which confirm that $A_S(2e^2/h)/G_2=1$ for different
density $n_L$ in the lower wire. (The two-terminal conductance $G_2$ has been normalized to $2e^2/h$.) Different samples correspond to different colors.}
\end{figure}
{\it It is important to keep in mind that this universal ratio constitutes a first proof of charge fractionalization in nonchiral Luttinger liquids (it certainly proves that the injected electron gives rise to two counterpropagating waves; however, it does not prove yet that the anomalous charges found are not the result of a quantum average; see Sec. 2.3)}. As a reminiscence of the free electron case, 
\begin{equation}
\frac{A_S(2 e^2/h)}{G_2} =  \frac{1}{g} \left(\frac{I_S^- - I_S^+}{I_S}\right) = 1,
\end{equation}
which has been recently confirmed experimentally \cite{Amir}; see Fig. 4. The value of the Luttinger parameter $g$ can be obtained directly from tunneling measurements as a function of energy (or $V_{SD}$). In the relatively weak-tunneling regime, this gives access to the electron spectral function, which shows two distinct peak features when $q$ is not too small to zero; see Fig. 5. From the Luttinger theory, those peaks are located at frequencies $\omega=u q$ and $\omega=v_Fq$ reflecting that the 
(forward) charge and spin excitations propagate at different velocities. Thus, from $ug = v_F$, one can directly extract the Luttinger exponent from tunnelings measurement as a function of energy (bias voltage $V_{SD}$) for different wavevectors $q$ or magnetic fields $B$.  For the experimental
setup of Ref. \cite{Auslaender,Amir}, one gets $0.4<g<0.5$ for the range of observed densities. This demonstrates that the system is far from the free electron regime.

Below, we discuss the parameter $\beta$ more thoroughly. 

\subsection{Maximal value of $\beta$ and perfect transmission}

The minimum value of $\beta$ corresponds to $\beta_{min}=0$.
Now, we are rather interested in determining the maximal value of $\beta$. 
\begin{figure}[ht]
\begin{center}
\vskip -2cm
\includegraphics[width=11.5cm,height=9cm]{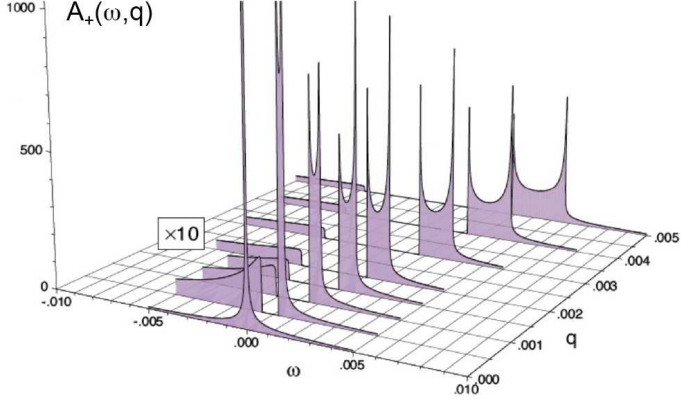}
\end{center}
\vskip -1.5cm
\caption{Electron spectral function in the Luttinger theory (for an electron from the right Fermi branch) as a function of frequency $\omega$ for different wavevectors $q$ measured from $k_F$ ($\Lambda/\hbar=1$ and 
$u=1$). Temperature is zero, $g=1/2$, and for simplicity, $L_F\rightarrow +\infty$ (a finite $L_F$ produces the broadening of the different peak structures). The $\omega<0$-part has been multiplied by 10 for clarity. Far from the Fermi ``surface'' (point), the electron spectral function reveals two peak features associated with the spin and right-moving charge mode (the counterpropagating charge mode also gives some spectral weight at negative $\omega$). One can determine $g$ from these two peaks.}
\end{figure}
For this purpose, let us set $\beta_1=\beta_3=\beta$ such that Eq. (\ref{G2}) becomes,
\begin{equation}
(V_1-V_3) = \left(\frac{2-\beta}{\beta}\right) (Y_1-W_1).
\end{equation}
The maximum value of
$\beta$ corresponds to the case where the current in the lower wire is equal to $(2e^2/h)(V_1-V_3)$; this describes perfect transmission where the upper left (right) wire and the lower wire form a unique wire that is perfectly coupled to the reservoir leads. This results in:
\begin{equation}
\frac{2ge^2}{h}(Y_1-W_1) = \frac{2ge^2}{h}\frac{\beta_{max}}{2-\beta_{max}} (V_1-V_3) = \frac{2e^2}{h}(V_1-V_3).
\end{equation} 
We infer that $\beta_{max}=2/(1+g)$. Therefore, we can write down:
\begin{equation}
0\leq \beta \leq \frac{2}{1+g}.
\end{equation}
{\it The asymmetry parameter $A_S$ and $G_2h/(2e^2)$ cannot exceed unity}. 

Below, we justify the maximal value $\beta_{max}=2/(1+g)$ from microscopic arguments. At perfect transmission, the upper left (right) wire and the lower wire form a unique wire\footnote{We assume that all the wires are described by the same Luttinger exponent.}, that is perfectly coupled to the reservoir leads. It is convenient to introduce the bare electron densities $\rho_{\pm}^e$ within the lower wire, such that
$\delta\rho=\rho_+ +\rho_- = \rho_+^e + \rho_-^e$, and \cite{Pham2}: 
\begin{equation}
\rho_+ = \frac{1+g}{2}\rho_+^e + \frac{1-g}{2}\rho_-^e \hskip 0.5cm \hbox{and} \hskip 0.5cm
\rho_- = \frac{1-g}{2}\rho_+^e + \frac{1+g}{2}\rho_-^e.
\end{equation}
Those equations are equivalent to Eqs. (B.2) and (B.3). The electron injection from the leads then induces the precise boundary conditions:
\begin{equation}
\frac{\partial H}{\partial \rho_{\pm}^e} = e V_{1,3}.
\end{equation}
Notice that an electron, injected from the contact O$_1$, is unambiguously converted into a right-moving electron at the left extremity $(x=0)$ of the system. 

Now, at the junction 1, we can write down the general equation: 
\begin{equation}
\frac{\partial H}{\partial \rho_{+}^e} = \frac{\partial H}{\partial \rho_+}\frac{1+g}{2} +
\frac{\partial H}{\partial \rho_-}\frac{1-g}{2}.
\end{equation}
At the junction 3, in a similar vein, we get:
\begin{equation}
\frac{\partial H}{\partial \rho_{-}^e} = \frac{\partial H}{\partial \rho_+}\frac{1-g}{2} +
\frac{\partial H}{\partial \rho_-}\frac{1+g}{2}.
\end{equation}

Using Eqs. (49) and (50), this allows us to verify that, 
\begin{equation}
V_1 = Y_1 \frac{1+g}{2} + W_1 \frac{1-g}{2} \hskip 0.5cm \hbox{and} \hskip 0.5cm
V_3 = Y_3 \frac{1-g}{2} + W_3 \frac{1+g}{2},
\end{equation}
which is indeed equivalent to $\beta_{max}=2/(1+g)$.

It is interesting to notice the analogy with the
case of a quantum wire ideally coupled to Fermi liquid leads at its extremities \cite{Ines,Ines2}. More precisely, if no charge is incident from the left lead (reservoir), resulting in $V_1=0$, we can check that $-\gamma=(g-1)/(g+1)=1-\beta_{max}$ is the reflection coefficient for a chiral ``quasiparticle'' incident on the contact $1$ $(3)$; $(1+\gamma)=\beta_{max}$ is the transmission coefficient for a chiral quasiparticle incident on a contact. Since $-\gamma<0$, a ``quasi-hole'' is reflected in analogy with Andreev reflection. Now, if no quasiparticle comes from the right at the contact $1$, this implies $Y_1=\beta_{max} V_1$ or in terms of currents, $(2ge^2/h)Y_1 = (\beta_{max} g) (2e^2/h)V_1$ where $(2e^2/h) V_1$ is the current stemming from the left reservoir. We deduce that the transmission coefficient for the incident flux from the left reservoir is $1-\gamma=2g/(g+1)=g\beta_{max}$. At perfect transmission, this allows us to check that
\cite{Ines}:
\begin{equation}
G_2 = \frac{2e^2}{h}(1-\gamma)(1+\gamma)\sum_{n=0}^{\infty} (-\gamma)^{2n} = \frac{2e^2}{h} 
\frac{g\beta_{max}^2}{1-(1-\beta_{max})^2} = \frac{2e^2}{h}.
\end{equation}
This allows us to check that the total charge which will be transmitted to the right lead at long times (or low frequency) is equal to the electron charge. A similar conclusion can be drawn when focusing on
the asymmetry parameter: $A_S=1$ when $\beta=\beta_{max}$, implying that $I_L=I_S$ and $I_R=0$.
A charge $e$ is transmitted from the source O$_2$ to the left contact O$_1$.
It should be noted that, in our context, $\beta_{max}=\beta_{max}(q\rightarrow 0)$ whereas for Fermi liquids connected at the extremities of a wire, $\beta_{max}=\beta_{max}(x=0)$ or $\beta_{max}=\beta_{max}(x=L)$.

\subsection{Weak-tunneling regime}

Now, we focus on the opposite weak-coupling regime $(\beta_i\ll 1)$ where the physics at the junctions $1$ and $3$ can be described by tunneling Hamiltonians; ${t}_1$ and ${t}_3$ denote the corresponding electron tunneling amplitudes or $T_1=|t_1|^2$ and $T_3=|t_3|^2$ denote the
electron tunneling probabilities. We can compute the (averaged) tunneling current at the left junction to lowest order in  ${t}_1$.

Using Appendix A and B.2,  for small $q=q_B-(k_F +k_{F2})\ll 1/L_F$, we obtain:
\begin{equation}
I_L(q) = \frac{2e^2}{h} |{t}_1|^2 \frac{(T/\Lambda)^{\nu}}{\Gamma(\nu+1)} \frac{{L}_F^2}{1+q^2{L}_F^2}\frac{1}{(\hbar v_F)^2} (W_1-V_1).
\end{equation}
The tunneling current at the junction $3$ takes the similar form:
\begin{equation}
I_R(q) = \frac{2e^2}{h} |{t}_3|^2 \frac{(T/\Lambda)^{\nu}}{\Gamma(\nu+1)}\frac{{L}_F^2}{1+q^2{L}_F^2} \frac{1}{(\hbar v_F)^2}(Y_3-V_3).
\end{equation}
The parameters $\beta_i$ can be justified by reminding the current conservation laws,
\begin{equation}
I_L = \frac{2ge^2}{h}(W_1-Y_1) = \frac{2ge^2}{h}\beta_1 (W_1-V_1),
\end{equation}
and,
\begin{equation}
I_R = \frac{2ge^2}{h}(Y_3-W_3) = \frac{2ge^2}{h} \beta_3 (Y_3-V_3).
\end{equation}
Finally, we identify:
\begin{equation}
\beta_i(q) =  \frac{1}{g} |t_i|^2  \frac{{L}_F^2}{1+q^2{L}_F^2} \frac{1}{(\hbar v_F)^2} 
\frac{(T/\Lambda)^{\nu}}{\Gamma(\nu+1)}.
\end{equation} 
The temperature dependence simply follows the product of the (momentum-resolved) electron tunneling densities of states in the two wires at low energy \cite{Karyn2,Fiete}. This result can be extended by including the (Coulomb) interaction between the wires \cite{Balents}; this only affects the form of the tunneling exponent $\nu$.

In the weak-tunneling limit, it should be noted that the prerequisite $\beta_1=\beta_3$ requires that 
$(V_1,V_3,V_{SD})\ll k_B T$ such that the parameters $\beta_i$ remain voltage-independent \cite{Balents,Karyn2}. In particular, remember that $\beta_1\gg \beta_3$ would imply $A_S\rightarrow 1$ independently of the charge fractionalization mechanism; consult Eq. (33).

\section{Conclusion}

When an electron is injected in the bulk of a quantum wire at a given Fermi point, say, $-k_F$, this gives rise to two counterpropagating pieces which carry charge $fe$ and $(1-f)e$, respectively, where $f=(1+g)/2$ \cite{Pham,Karyn,Ines,Imura,TMartin}. We have shown that the ratio between the asymmetry $A_S=(I_L-I_R)/I_S$ and the two-terminal conductance $G_2$ in the almost equilibrium regime, where the applied bias voltages are small compared to $k_B T/e$, allows to construct a novel dimensionless ratio reflecting the
charge fractionalization phenomenon,
\begin{equation}
\frac{A_S(2e^2/h)}{G_2} = \frac{(2f-1)}{g} = 1,
\end{equation}
in accordance with the recent experimental results \cite{Amir}. The actual values of $g$, and thus of $f$, were determined from tunneling measurements as a function of source-drain voltage $V_{SD}$ and $B$. The Luttinger theory predicts that those charges exist beyond the quantum and statistical average; see Sec. 2.3 and Appendix B.2 and B.3 \cite{Pham}. Finally, we have shown that in (gapless) Luttinger systems, it is possible to envision charge excitations with arbitrary charge; see Appendix B.4.  This should stimulate further challenging experiments.

We would like to acknowledge useful discussions with M. B\" uttiker, T. Giamarchi, Y. Oreg, K.-V. Pham, D. Prober, and I. Safi. We also thank E. Berg, E. Kim, and Y.  Oreg for
pointing out an error in an earlier version of Appendix B4.
Concerning this work, K.L.H. was supported
by Yale University, B.I.H. was supported in part by NSF grant DMR-0541988, and A.Y.
was supported by NSF grant DMR-0707484.

\appendix
\section{Current and Noise}

\subsection{Small Keldysh digest}

To be totally convinced by the charge fractionalization mechanism, one can compute the averaged current at a point $x$ and time $t$ in the lower wire.  Since the equalities $I_S^- = |\langle I(x\rightarrow 0)\rangle| = I_S f$ and $I_S^+ = \langle I(x\rightarrow L)\rangle$ should hold for any
bias voltage $V_{SD}$ applied between the upper and the lower wire, it is convenient to use the 
rigorous ``non-equilibrium'' Keldysh formalism. (Within the conventions used below, a right (left) current
is positive (negative).)

Here, we will apply the conventions of Ref. \cite{TMartin} and extend the analysis to our context where an electron is injected in the lower wire at a single Fermi point, say, $-k_F$. The current takes the general form,
\begin{equation}
\langle I(x,t)\rangle = \frac{1}{2}\sum_{\eta} \left\langle T_K\left[I(x,t_{\eta})e^{-\frac{i}{\hbar}\int_K dt' 
H_{t_2}(t')}\right]\right\rangle,
\end{equation}
where the coefficient $\eta=\pm$ identifies the upper/lower branch of the Keldysh contour $K$. Let us
first collect the lowest order contribution in the tunneling amplitude $t_2$. We get,
\begin{equation}
 \langle I(x,t)\rangle = -\frac{1}{4\hbar^2} \sum_{\eta\eta_1\eta_2} \eta_1 \eta_2 \left\langle T_K\left[ I(x,t_{\eta}) \int dt' dt'' H_{t_2}(t'_{\eta_1}) H_{t_2}(t''_{\eta_2})\right]\right\rangle.
 \end{equation}
The coefficients $\eta_{1,2}=\pm$ identify the upper/lower branch of the Keldysh contour. Since only
$\eta_1=-\eta_2$ will contribute, hereafter we fix $\eta_1=-\eta_2$.
Noting that $\partial_x F = \lim_{\zeta\rightarrow 0} (i\zeta)^{-1} \partial_x \exp(i\zeta F)$, we obtain:
\begin{eqnarray}
\hskip -0.5cm \langle I(x,t)\rangle &=& \frac{1}{4}\sqrt{\frac{2}{\pi}}ug |t_2|^2 \frac{e}{\hbar^2} \sum_{\epsilon\eta_1\eta\alpha} \int dt dt' \int dx' dx'' e^{i\epsilon \frac{e}{\hbar} V_{SD}(t'-t'')} e^{i\epsilon q(x'-x'')}
\\ \nonumber 
&\times& \left\langle T_K \Psi_{+2\alpha}^{-\epsilon}(x',t'_{\eta_1}) 
\Psi_{+2\alpha}^{\epsilon}(x'',t''_{-\eta_1})\right\rangle \\ \nonumber
&\times& \lim_{\zeta\rightarrow 0} (i\zeta)^{-1} \partial_x \langle T_K[e^{i\zeta\theta(x,t_{\eta})}\Psi_{-\alpha}^{\epsilon}(x',t'_{\eta_1})\Psi^{-\epsilon}_{-\alpha}(x'',t''_{-\eta_1})]\rangle.
\end{eqnarray}
The voltages are included through the Peierls substitution (as a phase in the electron creation/annihilation operator) and $\epsilon=\pm$ refers to a hole and electron, respectively.
It is convenient to introduce the compact notation $G_{-\alpha}^{\eta_1-\eta_1}(x'-x'',t'-t'') = \langle T_K \Psi_{-\alpha}^{\pm\epsilon}(x',t'_{\eta_1})\Psi^{\mp\epsilon}_{-\alpha}(x'',t''_{-\eta_1})\rangle$ and similarly we introduce the Green functions $G_{-2\alpha}^{\eta_1-\eta_1}$ for the electrons in the upper wire.  

A straightforward calculation leads to,
\begin{eqnarray}
\hskip -1cm \langle I(x)\rangle &=& \frac{1}{2}v_F |t_2|^2 \frac{e}{\hbar^2} \sum_{\eta_1\eta\alpha}  \int dx'  dx''
\cos(q(x'-x'')) \times \\ \nonumber
& & \int_{-\infty}^{+\infty} d{\tau} \sin\left(\frac{e V_{SD} \tau}{\hbar}\right) G_{-\alpha}^{\eta_1-\eta_1}(x'-x'',\tau)
G_{-2\alpha}^{\eta_1-\eta_1}(x'-x'',\tau) \times \\ \nonumber 
& & \int_{-\infty}^{+\infty} d\tau' \partial_x\left[G_{\theta\theta}^{\eta\eta_1}(x,x',\tau') - G_{\theta\theta}
^{\eta-\eta_1}(x,x'',\tau') + G_{\theta\phi}^{\eta\eta_1}(x,x',\tau') - G_{\theta\phi}
^{\eta-\eta_1}(x,x'',\tau')\right], 
\end{eqnarray}
where we have used that $ug=v_F$ for an ideal Luttinger liquid and where we have introduced the Keldysh Green's functions of the bosonic fields which have been thoroughly discussed in Ref. \cite{TMartin}. Now, we need to perform the integration over $\tau$. Since $|x'-x''|$ is cutoff by the length $L_F$ (see below Eq. (10)), at long times $t$, one can always simplify the electron Green's function as $G_{\alpha}(x'-x'',t) \propto \frac{1}{(t-i\epsilon)^{1+\kappa}}F[(x'-x'')/ut]$ where $F[(x'-x'')/ut]\rightarrow 1$ and $\kappa=-1/2+(g+g^{-1})/4=\nu/2$. The integration over $\tau$ gives a similar result as point-like tunnel coupling, and for temperatures $k_B T\gg eV_{SD}$ we find,
\begin{eqnarray}
\hskip -1cm \langle I(x)\rangle &=& i\frac{{L}_F}{2\hbar^2v_F}  |t_2|^2 \frac{e^2}{h}
\frac{(T/\Lambda)^{\nu}}{\Gamma(\nu+1)} V_{SD}\sum_{\eta_1\eta\alpha} \int_0^{{L}_F} dx
\cos(qx)\int_{-\infty}^{+\infty} d\tau' \sum_{\eta\eta_1} \eta_1 \\ \nonumber
&\times& \partial_x\left[G_{\theta\theta}^{\eta\eta_1}(x,x',\tau') - G_{\theta\theta}
^{\eta-\eta_1}(x,x'',\tau') + G_{\theta\phi}^{\eta\eta_1}(x,x',\tau') - G_{\theta\phi}
^{\eta-\eta_1}(x,x'',\tau')\right].
\end{eqnarray}
Here, we have assumed identical Fermi velocities in the two wires and $\Lambda$ is a high-temperature cutoff. Moreover, $\nu\geq 0$ is the usual tunneling exponent which obeys  $\nu=0$ for non-interacting electrons and which can be thoroughly evaluated by taking into account the Coulomb interaction between the wires; the general expression for $\nu$ can be found in Ref. \cite{Balents}. Ultimately, we obtain:
\begin{eqnarray}
-\int_{-\infty}^{+\infty} d\tau' \sum_{\eta\eta_1} \eta_1 \partial_x [G_{\theta\theta}^{\eta\eta_1}(x,x',\tau') - G_{\theta\theta}^{\eta-\eta_1}(x,x'',\tau') \\ \nonumber
+G_{\theta\phi}^{\eta\eta_1}(x,x',\tau') - G_{\theta\phi}^{\eta-\eta_1}(x,x'',\tau')]  \\ \nonumber
= -\frac{ig}{v_F} +\frac{i sgn(x-x')}{2v_F} + \frac{i sgn(x-x'')}{2v_F} .
\end{eqnarray}
At the left extremity of the lower wire, this expression results in:
\begin{eqnarray}
\langle I(x\rightarrow 0)\rangle = - \left(\frac{1+g}{2}\right) |t_2|^2 
\frac{2e^2}{h}\frac{1}{(\hbar v_F)^2}\frac{(T/\Lambda)^{\nu}}{\Gamma(\nu+1)} 
\frac{{L}_F^2}{1+q^2{L}_F^2} V_{SD}.
\end{eqnarray}
In contrast, at the right extremity of the lower wire, we obtain:
\begin{eqnarray}
\langle I(x\rightarrow L)\rangle = \left(\frac{1-g}{2}\right) |t_2|^2 
\frac{2e^2}{h}\frac{1}{(\hbar v_F)^2}\frac{(T/\Lambda)^{\nu}}{\Gamma(\nu+1)} 
\frac{{L}_F^2}{1+q^2{L}_F^2}V_{SD}.
\end{eqnarray}

Now, we need to evaluate the averaged tunneling current. This takes the form,
\begin{equation}
\langle I_S(t)\rangle = \frac{1}{2}\sum_{\eta} \left\langle T_K\left[I_S(t_{\eta})e^{-\frac{i}{\hbar}\int_K dt' H_{t_2}(t')}\right]\right\rangle,
\end{equation}
and $I_S(t)$ is given in Eq. (10). A similar calculation for $eV_{SD}\ll k_B T$ and $q\ll 1/{L}_F$ leads to:
\begin{eqnarray}
I_S(q,V_{SD}) &=& |t_2|^2 
\frac{e}{\hbar^2} \sum_{\eta} \int dx' \int dx'' \cos(q(x'-x'')) \\ \nonumber
&\times& \sum_{\eta\alpha} \int_{-\infty}^{+\infty} d\tau \sin\left(\frac{e V_{SD} \tau}{\hbar}\right) 
G_{-\alpha}^{\eta-\eta}(x'-x'',\tau)
G_{-2\alpha}^{\eta-\eta}(x'-x'',\tau) \\ \nonumber
&=& |t_2|^2 
\frac{2e^2}{h}\frac{1}{(\hbar v_F)^2}\frac{(T/\Lambda)^{\nu}}{\Gamma(\nu+1)} \frac{{L}_F^2}{1+q^2{L}_F^2} V_{SD}.
\end{eqnarray}
This expression is in accordance with Ref. \cite{Balents} and it reflects the spectral functions in the double-wire system \cite{Karyn2,Fiete}. This allows us to check that:
\begin{eqnarray}
|\langle I(x\rightarrow 0)\rangle| &=& I_S^- = \left(\frac{1+g}{2}\right) I_S(q\rightarrow 0,V_{SD})
\\ \nonumber
\langle I(x\rightarrow L)\rangle &=& I_S^+ = \left(\frac{1-g}{2}\right) I_S(q\rightarrow 0,V_{SD}).
\end{eqnarray}
From the definition of the current operator, the current $I(x)$ is oriented to the right.
The difference with the case of a point-like tunnel coupling should be noted; in that case,
the terms $G_{\theta\phi}$ in Eq. (A.5) would not be there and as a result $|\langle I(x\rightarrow 0)\rangle| = I_S/2$ and therefore on gets $\langle I(x\rightarrow L)\rangle = I_S/2$ \cite{TMartin}.

In the Keldysh formulation, the current noise takes the form:
\begin{equation}
 S(x,t,x',t') = -\frac{1}{4\hbar^2} \sum_{\eta\eta_1\eta_2} \eta_1 \eta_2 \left\langle T_K\left[ I(x,t_{\eta}) I(x',t'_{-\eta})\int dt'' dt''' H_{t_2}(t''_{\eta_1}) H_{t_2}(t'''_{\eta_2})\right]\right\rangle.
 \end{equation}
 The calculation is essentially similar as the one for the computation of the averaged current. 
 This involves the computation of:
  \begin{equation}
 \lim_{\zeta\rightarrow 0} (\zeta)^{-2} \partial_x \partial_{x'} \langle T_K[e^{i\zeta\theta(x,t_{\eta})}
 e^{-i\zeta\theta(x',t'_{-\eta})}\Psi_{-\alpha}^{\epsilon}(x',t''_{\eta_1})\Psi^{-\epsilon}_{-\alpha}(x'',t'''_{-\eta_1})]\rangle.
\end{equation}
Assuming that $q\rightarrow 0$, for the zero-frequency noise $S(x,x',\omega=0)$, we find :
\begin{eqnarray}
S(0,0,\omega=0) &=& 2 \left(\frac{1+g}{2}\right)e |\langle I(x\rightarrow 0)\rangle| \\ \nonumber
  S(L,L,\omega=0) &=& 2 \left(\frac{1-g}{2}\right)e \langle I(x\rightarrow L)\rangle.
  \end{eqnarray}
  
 \subsection{Linear response in $t_2$}

Here, we apply  the ``equilibrium'' interaction representation to compute the current $I(x,t)$. We note
$|0\rangle$ the ground state of the Luttinger theory.

By definition, one gets:
\begin{equation}
\langle I(x,t)\rangle = \sqrt{\frac{2}{\pi}}uge \langle 0| T\left[\partial_x\theta(x,t) e^{-\frac{i}{\hbar}
\int_{-\infty}^{+\infty} dt' H_{t_2}(t')}\right] |0\rangle.
\end{equation}
Firstly, we compute the averaged current to lowest order in the electron tunneling amplitude $t_2$ using
that $\partial_x \theta= \frac{1}{2}(\partial_x\theta_+ +\partial_x\theta_-)$.
One can also use that $\partial_x F = \lim_{\zeta\rightarrow 0} (i\zeta)^{-1} \partial_x \exp(i\zeta F)$.
Therefore, the calculation of $\langle I(x,t)\rangle$ can be decomposed into two (chiral) parts. The first part involves $\partial_x\theta_-$:
\begin{eqnarray}
 \langle I_-(x,t)\rangle &=& -\sqrt{\frac{2}{\pi}}ug |t_2|^2 \frac{e}{\hbar^2}\int_{-\infty}^{t} dt' \int_{-\infty}^{t'} dt'' \int dx' \int dx'' \\ \nonumber
&\times& e^{i\frac{e}{\hbar}V_{SD}(t'-t'')} e^{iq(x' - x'')}\sum_{\alpha=\uparrow,\downarrow} \langle \Psi_{+2\alpha}^{\dagger}(x',t') \Psi_{+2\alpha}(x'',t'')\rangle \\ \nonumber
&\times& \lim_{\zeta\rightarrow 0} (i\zeta)^{-1} \partial_x \langle T[e^{i\zeta\theta_-(x,t)}\Psi_{-\alpha}(x',t')\Psi^{\dagger}_{-\alpha}(x'',t'')]\rangle.
\end{eqnarray}
We find,
\begin{eqnarray}
&&\lim_{\zeta\rightarrow 0} (i\zeta)^{-1} \partial_x \langle T[e^{i\zeta\theta_{\mp}(x,t)}
\Psi_{-\alpha}(x',t')\Psi^{\dagger}_{-\alpha}(x'',t'')]\rangle \\ \nonumber
&=& \langle \Psi_{-\alpha}(x',t')\Psi^{\dagger}_{-\alpha}(x'',t'')\rangle 
\times  i\left(\frac{1\pm g}{2}\right)\sqrt{\frac{\pi}{2}} \partial_x\langle \theta_{\mp}(x,t)\theta_{\mp}(x',t')\rangle.
\end{eqnarray}
Therefore,
\begin{eqnarray}
\hskip -0.5cm \langle I_-(x,t)\rangle &=& -i\left(\frac{1+g}{2}\right)|t_2|^2\frac{e}{\hbar^2}ug \int_{-\infty}^{t} dt' \int_{-\infty}^{0} dt'' \int dx' dx'' e^{-i\frac{e}{\hbar}V_{SD}t''} \\ \nonumber
&\times& e^{iq(x' - x'')} \sum_{\alpha=\uparrow,\downarrow}
\langle \Psi_{+2\alpha}^{\dagger}(x',0) \Psi_{+2\alpha}(x'',t'')\rangle 
\langle\Psi_{-\alpha}(x',0)\Psi^{\dagger}_{-\alpha}(x'',t'')\rangle \\ \nonumber
&\times& \partial_x\langle \theta_-(x,t)\theta_-(x',t')\rangle.
\end{eqnarray}
Now, at relatively small temperatures, we can use:
\begin{eqnarray}
\langle \theta_{\pm}(x,t)\theta_{\pm}(0,0) - \theta_{\pm}(x,t)^2\rangle = \frac{1}{g\pi} \ln\left(\frac{a/u}{i((t-i\epsilon)\mp \frac{x}{u})}\right),
\end{eqnarray}
and,
\begin{equation}
\partial_x \langle \theta_{\pm}(x,t)\theta_{\pm}(0,0) - \theta_{\pm}(x,t)^2\rangle = \partial_x
 \langle \theta_{\pm}(x,t)\theta_{\pm}(0,0) \rangle,
 \end{equation}
such that:
\begin{eqnarray}
\langle I_-(x,t)\rangle &=&  -\left(\frac{1+g}{2}\right) |t_2|^2\frac{e}{\hbar^2} \int_{-\infty}^{t} dt'
 \int dx' dx'' \int_{-\infty}^{0} dt'' \\ \nonumber
 &\times& \delta(t'-t +(x'-x)/u) e^{-i\frac{e}{\hbar}V_{SD}t''} e^{iq(x' - x'')} \\ \nonumber
 &\times& \sum_{\alpha=\uparrow,\downarrow}
 \langle \Psi_{+2\alpha}^{\dagger}(x',0) \Psi_{+2\alpha}(x'',t'')\rangle 
\langle\Psi_{-\alpha}(x',0)\Psi^{\dagger}_{-\alpha}(x'',t'')\rangle.
\end{eqnarray}
This contribution is non-zero only if $(x',x'')>x$; essentially, $I_-$ is a left-going current. At the left extremity of the lower wire, this expression gives:
\begin{eqnarray}
\hskip -1cm \langle I_-(x\rightarrow 0)\rangle = -\left(\frac{1+g}{2}\right)|t_2|^2\frac{e}{\hbar^2}\int_{-\infty}^{0} dt'' \int dx' \int dx'' e^{-i\frac{e}{\hbar}V_{SD}t''} e^{iq(x' - x'')} \\ \nonumber
\times  \sum_{\alpha=\uparrow,\downarrow}  \langle \Psi_{+2\alpha}^{\dagger}(x',0) \Psi_{+2\alpha}(x'',t'')\rangle 
\langle \Psi_{-\alpha}(x',0)\Psi^{\dagger}_{-\alpha}(x'',t'')\rangle.
\end{eqnarray}
Finally, we can evaluate the averaged tunneling current,
\begin{equation}
\langle I_S(t)\rangle = \langle 0|T[I_S(t)
\exp(-\frac{i}{\hbar}\int_{-\infty}^{+\infty} dt' H_{t_2}(t'))]|0\rangle.
\end{equation}
 to second order in $t_2$. This allows
us to check that:
\begin{equation}
|\langle I_-(x\rightarrow 0)\rangle| = I_S^- = \left(\frac{1+g}{2}\right) I_S(q\rightarrow 0,V_{SD}).
\end{equation}
The second part of the current involving $\partial_x\theta_+(x,t)$ can be treated in the same way;
it gives a finite contribution only if $x$ is located at the right extremity of the lower wire, and we
identify:
\begin{equation}
\langle I_+(x\rightarrow L)\rangle = I_S^+ = \left(\frac{1-g}{2}\right) I_S(q\rightarrow 0,V_{SD}).
\end{equation}

\section{Note on Charge Fractionalization}

\subsection{Universal Ratio}

In the case of unidirectional injection, the ratio 
\begin{equation}
\frac{I_S^- - I_S^+}{I_S} = (2f-1)=g,
\end{equation}
examplifies the charge fractionalization in a Luttinger liquid. In fact, one can always write the relations (this point has also been noted in Ref. \cite{Imura}),
\begin{equation}
I_S^- = \frac{1+g}{2}I_{S,-k_F} +\frac{1-g}{2}I_{S,+k_F}, 
\end{equation}
and,
\begin{equation}
I_S^+ = \frac{1-g}{2} I_{S,-k_F}+\frac{1+g}{2}I_{S,+k_F},
\end{equation}
where $I_{S,-k_F}$ and $I_{S,+k_F}$ are the injected currents at the two Fermi points in the lower wire. 
We argue that momentum-resolved tunneling is crucial to observe charge fractionalization since it satisfies $I_{S,-k_F}=I_S$ and $I_{S,+k_F}=0$; this corresponds to a unidirectional electron injection. In contrast,  for point-like tunnel coupling, one finds $I_{S,+k_F}=I_{S,-k_F}=I_S/2$ resulting in $I_S^-=I_S^+=I_S/2$ \cite{TMartin}. This case is not judicious for demonstrating charge fractionalization \cite{TMartin}.

Assuming that the Luttinger theory is applicable, this ratio remains valid for spinless electrons and for multi-band systems such as carbon nanotubes. 

For spinless electrons, {\it i.e.} for a strong Zeeman effect, one gets \cite{Pham},
\begin{equation}
\rho_{\pm} = \frac{1}{2} \sqrt{\frac{1}{\pi}}\left[-\partial_x \phi \pm g\partial_x \theta\right],
\end{equation}
and,
\begin{equation}
H =  \int_0^L dx \left(\frac{u \hbar \pi}{g}\left[\rho_+^2 +\rho_-^2\right]\right),
\end{equation}
The electron operator turns into \cite{Pham},
\begin{equation}
\Psi^{\dagger}_{-}(x) = \frac{1}{\sqrt{2\pi a}}\exp \left[-i\sqrt{\pi}\left(\theta(x) +\phi(x)\right)\right],
\end{equation}
which immediately restores the equations (\ref{IS-}) and (\ref{IS+}). Those equations are
also applicable in the low-density regime which is characterized by completely spin-incoherent excitations \cite{Greg}. On the other hand, carbon nanotubes are
described by a two-band model whose fermionic operators obey,
\begin{equation}
\Psi_{-i\alpha}^{\dagger} = \frac{1}{\sqrt{2\pi a}} e^{-i\sqrt{\pi}(\theta_{i\alpha} + \phi_{i\alpha})},
\end{equation}
and the lowerscript $i=1,2$ refers to the two bands. In (metallic) carbon nanotubes, the current
is related to the total charge density $\rho = \sum_{i\alpha} \partial_x \phi_{i\alpha}/\sqrt{\pi} = 2\partial_x\phi/\sqrt{\pi}$ which is described by a Luttinger theory (the three other neutral modes are described by a free fermion model). Now, by noting the correspondence between the field $\theta_{i\alpha}$ and the total superfluid phase $\theta$, one obtains,
\begin{equation}
\left[ \Psi^{\dagger}_{-i\alpha}(x'), \rho_{\mp}(x)\right] = -\frac{1\pm g}{2} \Psi^{\dagger}_{\mp i\alpha}(x')\delta(x-x').
\end{equation}
We immediately deduce that, by injecting an electron in a metallic (armchair) carbon nanotube at one Fermi point (only), this would produce the same irrational charge quantum numbers \cite{TMartin}. 

\subsection{Chiral Eigenstates of the purely linear dispersion} 

The chiral operators associated with the chiral charge sectors obey \cite{Pham},
\begin{equation}
{L}_{\pm}^{{N}_{\pm}}(x,t) = \exp-i\sqrt{\frac{\pi}{2}}{N}_{\pm}\Theta_{\pm}(x,t),
\end{equation}
where
\begin{equation}
N_{\pm} = \frac{N\pm gJ}{2},
\end{equation}
and the fields $\sqrt{\pi/2}\Theta_{\pm} = \sqrt{\pi/2}(\theta \mp \phi/g)$ are conjugate to the densities 
$\rho_{\pm}$. Assuming a perfect linear spectrum, one can show that their Fourier transforms are exact eigenstates of the Luttinger Hamiltonian (consult Appendix A of Ref. \cite{Pham}).

It is relevant to observe that the chiral chemical potentials $W$ and $Y$ can be absorbed through the
re-definition:
\begin{equation}
\sqrt{\frac{\pi}{2}} \Theta_{\pm} \rightarrow \sqrt{\frac{\pi}{2}} \Theta_{\pm} +\frac{e}{\hbar} (Y,W)t.
\end{equation}
The electron operator takes the precise form \cite{Karyn,Karyn2},
\begin{equation}
\Psi^{\dagger}_{-\alpha}(x) = \frac{e^{i\chi t}}{\sqrt{2\pi a}}L_-^{\frac{1+g}{2}}L_+^{\frac{1-g}{2}}S_{-\alpha}(x),
\end{equation}
where,
\begin{equation}
\chi = \frac{e}{\hbar}\left(\frac{1+g}{2}W + \frac{1-g}{2}Y\right).
\end{equation}
When the tunnel coupling between the lower wire and the left and right upper wires is
weak, this implies that $W = Y +{O}(\beta)$, and we can safely approximate:
\begin{equation}
\chi \approx  \frac{e}{\hbar} W \approx \frac{e}{\hbar} Y.
\end{equation}

Eq. (B.12) gives a physical understanding to the Dzyaloshinskii-Larkin Green's
function \cite{Larkin} and of the form of the electron spectral function in Fig. 5.

\subsection{Note on Dispersion Nonlinearity}

The exact solvability of the Luttinger model relies on the assumption of strictly linear dispersion relation. 
Here, we briefly discuss the effects of the dispersion nonlinearity on the chiral phonons. The phonon
operators are defined as,
\begin{equation}
b_q = \sqrt{\frac{g|q|}{2}}\left(\theta_q - \frac{q}{g|q|}\phi_q\right),
\end{equation}
and,
\begin{equation}
b_q^{\dagger} = \sqrt{\frac{g|q|}{2}}\left(\theta_{-q} - \frac{q}{g|q|}\phi_{-q}\right),
\end{equation}
where $\theta_q$ and $\phi_q$ are the Fourier transforms of $\theta(x)$ and $\phi(x)$, respectively. The
phonons modes are also related to the chiral modes via,
\begin{equation}
b_{q>0} = \sqrt{\frac{g|q|}{2}}\Theta_{+,q} \hskip 0.5cm \hbox{and} \hskip 0.5cm
b_{q<0} = \sqrt{\frac{g|q|}{2}}\Theta_{-,q},
\end{equation}
where the modes $\Theta_{\pm}$ have been defined in Eq. (B.11). In the Luttinger theory, the charge Hamiltonian can be precisely re-written as $H=H_+ + H_-$, and the chiral charge Hamiltonians take the precise forms (for simplicity, from now on, we consider spinless fermions),
\begin{equation}
H_{\pm} = \sum_{\pm q} \hbar u|q| : b^{\dagger}_q b_q: + \frac{\hbar \pi u}{L g}\left(\frac{\hat{N} \pm g \hat{J}}{2}\right)^2,
\end{equation}
where $\hat{N}$ and $\hat{J}$ are operators such that $\langle \hat{N}\rangle = N$ and $\langle \hat{J}\rangle = J$. Now, let us include a non-linear dispersion and approximate the
electron spectrum:
\begin{equation}
\xi_p^{\pm} = \pm v_F p +\frac{p^2}{2m}\hskip 1cm \hbox{where} \hskip 1cm p=k\mp k_F.
\end{equation}
The presence of the finite mass $m$ breaks the particle-hole symmetry and may affect the spectral function in the electron and hole regions in different ways \cite{Khodas}. On the other hand, one
expects that the Luttinger bosons $b^{\dagger}_{q}$ are not eigenstates anymore of the system.
More precisely, one can compute the damping rate of the chiral phonons (or chiral eigenstates).

The non-linear dispersion produces triple collisions \cite{Samokhin}:
\begin{eqnarray}
\delta H = \int \frac{d k_1}{2\pi} \frac{d k_2}{2\pi} \frac{d k_3}{2\pi} V(k_1,k_2,k_3)
(b_{k_1} b^{\dagger}_{k_2} b^{\dagger}_{k_3} \delta(k_1-k_2-k_3) \\ \nonumber
+ (\hbox{all permutations of}\  k_1, k_2, k_3)) \\ \nonumber
+ V(k_1,k_2,k_3)(b_{k_1} b_{k_2} b_{k_3} \delta(k_1+k_2+k_3) +h.c.),
\end{eqnarray}
where,
\begin{equation}
V(k_1,k_2,k_3) = \frac{1}{6m} \frac{k_1 k_2 k_3}{|k_1 k_2 k_3|^{1/2}}(\prod_i (\theta(k_i)\cosh\varphi
- \theta(-k_i)\sinh \varphi) - (k_i\rightarrow -k_i)),
\end{equation}
and,
\begin{equation}
e^{2\varphi} = g.
\end{equation}
The self-energy function $\Sigma$ can be computed in a self-consistent way following Ref. \cite{Samokhin}. The damping rate for $\omega=u|p|\rightarrow 0$ takes the form,
\begin{equation}
\Gamma (T \neq 0) \propto  |\lambda| \sqrt{T} |p|^{3/2},
\end{equation} 
and $\lambda\rightarrow 1/(2\pi m)$ in the limit of vanishing electron-electron interactions. Remember that, in the limit $\hbar u|p|=\hbar\omega\ll k_B T$, the damping rate of Luttinger bosons varies as $\sqrt{T}$ and $p^{3/2}$. The long-wavelength bosonic excitations obey the same decay rate as the sound attenuation in 1D classical liquids. Assuming that $m$ is finite, the Luttinger bosons acquire a finite lifetime and the results found in this paper should remain valid as long as,
\begin{equation}
u|p| = \hbar \omega \geq |\lambda| \sqrt{T} |p|^{3/2} \rightarrow |p| \leq \frac{u^2}{\lambda^2 |T|},
\end{equation}
{\it i.e.}, for small applied bias voltages (energy).

\subsection{Production of an excitation with arbitrary charge in a Luttinger liquid}

Here we show how, in principle, one can inject a pulse with arbitrary definite charge into an ideal Luttinger liquid.

Consider an ideal {\it infinite}  one-dimensional wire, free of scatterers, which has one value, $g_1$,  for the Luttinger parameter in the region $ x \ll 0$,  and a different value $g_2$ in the region 
$0 \ll x $. We assume that in the vicinity of the origin  the value of $g$ varies ``adiabatically" between $g_1$ and $g_2$, over a distance scale $a$ which is very large compared to the inverse of the Fermi momentum $k_F$.  In this case, we may consider that the total momentum of the system is conserved, as the junction cannot scatter electrons  from $k_F$ to $-k_F$.

Suppose that at  time $t=0$, at a point $x = -b\ll 0$, {\it far} to the left of the origin, we inject a right-moving electron to into the one-dimensional wire, via a momentum-conserving tunnel junction.  
This will produce a right-moving excitation with charge $f_1 e$ and a left-moving excitation with charge $(1-f_1)e$, where $f_1 =( 1+ g_1)/2$.  The right-moving excitation will reach the interface at $x=0$ 
after a time $t \approx  b/u_1$, where $u_1$ is the charge velocity in the region $x<0$. This will produce
a right-moving charge $N_+ e$ going into the  region $x>0$ with a Luttinger parameter $g_2$ (and a plasmon velocity $u_2=v_F/g_2$) as well as
a second left-moving ``reflected'' charge $N_- e$, determined by the  conservation laws for charge and momentum:
\begin{equation}
N_+ + N_- = f_1, \hskip 1cm u_2 N_+ - u_1 N_- = f_1 u_1.
\end{equation} 
Solving those equations, we find:
\begin{equation}
N_- = f_1 \frac{g_1 - g_2}{g_1+g_2}, \hskip 1cm  N_+ = f_1 \frac{2g_2} {g_1+g_2}.
\label{npm}
\end{equation}
Thus, the incident chiral quasiparticle has a reflection coefficient $-\gamma=(g_1-g_2)/(g_1+g_2)$, in agreement with the results by Safi and Schulz \cite{Ines} and those of Sec. 3.3.

As  $g_1$  can be chosen arbitrarily in the interval $0 < g_1  \le 1$,  the size of the right-moving charge is not determined by the  Luttinger parameter  $g_2$ for the region $x>0$, but can have an arbitrary value  in the range  
\begin{equation}
\frac{2g_2}{g_2+1} \leq N_+ <  1 .
\label{ineq1}
\end{equation}
Similarly, the  left-moving reflected charge, can take on arbitrary values in the range 
\begin{equation}
f_1 \frac{g_1 - 1}{g_1+1} \leq N_- < f_1 .
\label{ineq2}
\end{equation}

In the case where the injection point $-b$ is close to the origin, the right-moving transmitted charge will have the same value $N_+ e$ as before, given by Eq. (\ref{npm}).  However, 
the left-moving reflected charge $N_-$ will merge physically with the original left-moving pulse, of strength $(1-f_1)$, to give a single left-moving pulse of charge $(1-N_+)e$.

We may also consider the case where we initially inject a {\em left-moving} electron into the wire at $x= -b$, thus producing an initial right-moving pulse of strength $1-f_1$ rather than $f_1$.  Now we must replace $f_1$ by $1-f_1$ in the formula for $N_+$. Then, by varying $g_1$ we can get any value of $N_+$ in the range $0<N_+ \leq 1$.  Arbitrary values of $N_+ $ can be obtained by injecting multiple electrons or holes at $x=-b$.

This scheme may also be used, in principle, to produce an excitation of arbitrary charge along the edge of a fractional quantized Hall system. To see this, let us suppose that the system is completely spin-polarized, and suppose that the Luttinger liquid in the region $x>0$ is a narrow wire with interaction parameter $g_2=1/3$.   At some point $x_0$, far to the right of the origin, we allow the wire to widen out into a broad strip containing a  two-dimensional electron gas in a magnetic field, at Landau-level filling fraction $\nu=1/3$.
In principle, this can be done adiabatically, by careful control of the electron density and the lateral confining potential at each point, and it should be possible to keep the Luttinger parameter fixed at $1/3$ throughout the transition region \cite{Chklov}.  Now, with an appropriate choice of $g_1$ in the region $x<0$, and by injecting  left-moving or right-moving electrons at the point $x=-b$, we can produce an arbitrary charge $N_+$ moving along the edge of the quantized Hall strip.

\end{document}